\documentstyle[12pt]{article}

\font\twlgot =eufm10 scaled \magstep1
\font\egtgot =eufm8
\font\sevgot =eufm7
\font\twlmsb =msbm10 scaled \magstep1
\font\egtmsb =msbm8
\font\sevmsb =msbm7

\newfam\gotfam

\textfont\gotfam\twlgot
\scriptfont\gotfam\egtgot
\scriptscriptfont\gotfam\sevgot

\newfam\msbfam
\textfont\msbfam\twlmsb
\scriptfont\msbfam\egtmsb
\scriptscriptfont\msbfam\sevmsb
\def\Bbb{\protect\pBbb}
\def\pBbb{\relax\ifmmode\expandafter\Bb\else\typeout{You cann't use
Bbb in text mode}\fi}
\def\Bb #1{{\fam\msbfam\relax#1}}

\def\thebibliography#1{\bigskip\section*{}\bigskip\list
{$^{\arabic{enumi}}$}{\settowidth\labelwidth{#1}\leftmargin\labelwidth
\advance\leftmargin\labelsep
\usecounter{enumi}}
\def\newblock{\hskip .11em plus .33em minus .07em}
\sloppy\clubpenalty4000\widowpenalty4000
\sfcode`\.=1000\relax}

\def\op#1{\mathop{\fam0 #1}\limits}

\newcommand{\beq}{\begin{equation}}
\newcommand{\eeq}{\end{equation}}
\newcommand{\ben}{\begin{eqnarray}}
\newcommand{\een}{\end{eqnarray}}
\newcommand{\be}{\begin{eqnarray*}}
\newcommand{\ee}{\end{eqnarray*}}
\newcommand{\bea}{\begin{eqalph}}
\newcommand{\eea}{\end{eqalph}}
\newcommand{\cA}{{\cal A}}

\newcommand{\cH}{{\cal H}}

\newcommand{\bL}{{\bf L}}

\newcommand{\bH}{{\bf H}}

\newcommand{\al}{\alpha}

\newcommand{\bt}{\beta}
\newcommand{\dl}{\delta}
\newcommand{\la}{\lambda}
\newcommand{\La}{\Lambda}
\newcommand{\f}{\phi}

\newcommand{\Om}{\Omega}

\newcommand{\g}{\gamma}
\newcommand{\G}{\Gamma}

\newcommand{\vt}{\vartheta}
\newcommand{\vf}{\varphi}

\newcommand{\lng}{\langle}
\newcommand{\rng}{\rangle}

\newcommand{\si}{\sigma}
\newcommand{\Si}{\Sigma}
\newcommand{\w}{\wedge}
\newcommand{\wt}{\widetilde}
\newcommand{\wh}{\widehat}
\newcommand{\ol}{\overline}
\newcommand{\dr}{\partial}

\newcommand{\ot}{\otimes}

\let\ssection=\section
\renewcommand{\section}{\setcounter{equation}{0}\ssection}

\newcounter{eqalph}
\newcounter{equationa}
\newcounter{remark}
\newcounter{example}
\newcounter{theorem}
\newcounter{proposition}
\newcounter{lemma}
\newcounter{corollary}
\newcounter{definition}
\newenvironment{eqalph}{\stepcounter{equation}
\setcounter{equationa}{\value{equation}}
\setcounter{equation}{0}

\begin{eqnarray}}{\end{eqnarray}\setcounter{equation}{\value{equationa}}}

\def\therexample{\arabic{remark}}
\def\thetheorem{\arabic{theorem}}

\def\thedefinition{\arabic{definition}}

\newcommand{\mar}[1]{}

\begin{document}
\hbox{}

\thispagestyle{empty}

{\parindent=0pt

{\large\bf Nonadiabatic holonomy operators in classical and
quantum completely integrable systems}
\bigskip

{\sc G.Giachetta}\footnote{Electronic mail: giovanni.giachetta@unicam.it}

{\sl Department of Mathematics and Informatics, University of Camerino,
62032 Camerino (MC), Italy}

\medskip

{\sc L.Mangiarotti}\footnote{Electronic mail: luigi.mangiarotti@unicam.it},

{\sl Department of Mathematics and Informatics, University of Camerino,
62032 Camerino (MC), Italy}

\medskip

{\sc G. Sardanashvily}\footnote{Electronic mail:
sard@grav.phys.msu.su}

{\sl Department of Theoretical Physics,
Moscow State University, 117234 Moscow, Russia}

\bigskip

Given a completely integrable system, 
we associate to any connection
on a fiber bundle in invariant tori over a parameter manifold 
the classical and quantum holonomy operator (generalized 
Berry's phase factor), without any adiabatic approximation.

}

\bigskip
\bigskip

\noindent
{\bf I. INTRODUCTION}
\bigskip

At present, holonomy operators in quantum systems attract special
attention in connection with
quantum computation (see, e.g., Refs. [1-3]). They exemplify the
non-Abelian generalization of 
Berry's geometric phase by means of driving a finite level
degenerate eigenstate of a Hamiltonian over a parameter manifold.
The key point is that
a geometric phase depends only on the geometry of a path executed  and,
therefore, provides a possibility to perform quantum gate operations in
an intrinsically fault-tolerant way. The problem lies in separation of a 
geometric phase factor from the total evolution operator without using an
adiabatic assumption. Firstly, holonomy quantum computation implies
exact cyclic evolution, but exact adiabatic cyclic evolution almost
never exists. Secondly, an adiabatic condition requires that the
evolution time must be long enough. 

A nonadiabatic Abelian phase was 
discovered by Aharonov and Anandan who considered a loop in a projective
Hilbert space instead of a parameter space.$^4$ Non-Abelian
generalization of the Aharonov--Anandan phase has been studied under
rather particular assumption.$^5$. Moreover, a non-Abelian
Aharonov--Anandan phase fail to be separated from the dynamic one 
in general. Recently, several schemes using the
Aharonov--Anandan phase were proposed for nonadiabatic geometric
gates.$^{6-8}$ 

In a general setting, let us consider a linear 
(not necessarily finite-dimensional) dynamical system 
$\dr_t\psi=\wh S \psi$ whose  linear (time-dependent) 
dynamic operator $\wh S$ falls into the sum
\mar{r0}\beq
\wh S=\wh S_0 + \Delta=\wh S_0 +\Delta_\al\dr_t\xi^\al, \label{r0}
\eeq
where $\xi(t)$ is a function of time
taking its values in a finite-dimensional smooth real parameter
manifold $\Si$ coordinated by $(\si^\al)$. Let assume that:
(i) the operators $\wh S_0(t)$ and $\Delta(t')$ commute for all
instants $t$ and $t'$, and (ii) the operator $\Delta$ 
depends on time only through $\xi(t)$. 
Then the evolution operator $U(t)$ can be represented by the product of
time-ordered exponentials
\mar{rr}\beq
U(t)=U_0(t)\circ U_1(t) = T\exp\left[\op\int^t_0 \wh S_0dt'\right]
\circ T\exp\left[\op\int^t_0 \Delta dt'\right], \label{rr}
\eeq
where the second one is brought into the ordered exponential
\mar{r4}\beq
U_1(t) = T\exp\left[\op\int^t_0 \Delta_\al(\xi(t'))\dr_t\xi^\al(t') dt'\right]
=T\exp\left[\op\int_{\xi[0,t]} \Delta_\al(\si)d\si^\al\right] \label{r4}
\eeq 
along the curve $\xi[0,t]$ in the parameter manifold $\Si$. 
It is a nonadiabatic geometric factor depending only on the trajectory
of the parameter function $\xi$. Accordingly, $\Delta$ is a holonomy
operator. The geometric factor
(\ref{r4}) is well defined if 
$\Delta_\al d\si^\al$
is an Ehresmann connection on a fiber bundle over a 
parameter manifold $\Si$. Then this factor  
is a displacement operator along an arbitrary curve $\xi[0,t]\subset \Si$. 

A problem is that the above mentioned commutativity condition (i) 
is very restrictive.
Moreover, it need not be preserved under time-dependent transformations.

For instance, let us consider a Hamiltonian system of dynamic variables
$(q,p)$. Written with
respect the initial data coordinates $(q_0,p_0)$, its Hamiltonian 
$\cH(q_0,p_0)$ vanishes. Given these coordinates 
$(q_0,p_0)$, let one can introduce a perturbed Hamiltonian 
$\cH_\xi(q_0,p_0,\xi(t))$ which
depends on parameter functions $\xi(t)$ and generates a holonomy operator
$\Delta$ (\ref{r0}). Then the evolution operator of the perturbed
Hamiltonian system reduces to the geometric factor (\ref{r4}).
Relative to the original variables $(q,p)$, a
Hamiltonian of this perturbed Hamiltonian system is 
\be
\cH'=\cH(q,p,t) + \cH_\xi(q_0(t,q,p),p(t,q,p),\xi(t)).
\ee
However, the corresponding evolution operator does not fall into the product
(\ref{rr}) because a Hamiltonian $\cH$ is not a function under time-dependent 
transformations and, consequently, the Poisson bracket $\{\cH,\cH_\xi\}$
with respect to original variables $(q,p)$ need not vanish.  

Nevertheless, basing on this example, we can essentially extend the
class of dynamical systems admiting a nonadiabatic geometric phase.
We aim to describe dynamical systems where the commutativity condition (i)
is not satisfied, but a part of dynamic variables is driven only by a holonomy 
operator. These are completely integrable Hamiltonian systems. 

Let us consider a completely integrable Hamiltonian system (henceforth
CIS) of $m$ degrees of freedom around its invariant tori $T^m$. 
We show that, being
constant under an internal evolution, its action
variables are driven 
only by a perturbation holonomy operator $\Delta$ which can be
associated to an arbitrary  
connection on a fiber bundle 
\mar{r25}\beq
\Si\times T^m\to \Si. \label{r25}
\eeq
This
holonomy operator is defined
with respect to the initial data action-angle 
coordinates without any adiabatic approximation.
Then we return to the original action-angle coordinates.
The key point is that both classical evolution
of action variables and mean values of quantum action operators
relative to original action-angle coordinates are determined 
in full by the dynamics of initial data action and 
angle variables.
 
The plan of the paper is as follows. 
Section II addresses classical 
time-dependent CIS. The key point is that
any time-dependent CIS of $m$ degrees of freedom is extended 
to an autonomous CIS of $m+1$ degrees of freedom$^{9-11}$
and, as a consequence,
can be provided with action-angle variables around a regular 
instantly compact invariant manifold.$^{10,11}$

In Section III, we introduce the holonomy operator in a 
classical CIS by use of
the fact that a generic Hamiltonian of a mechanical system 
with time-dependent parameters contains a term which is linear
both in momenta and the temporal derivative of a parameter
function.$^{12,13}$   
This term comes from a connection on the configuration 
space of the system fibered over a parameter manifold.

Section IV is devoted to geometric quantization  
of a CIS with respect to the angle polarization.
This polarization leads to the Schr\"odinger representation of action
variables  
in the separable Hilbert
space of smooth complex functions on $T^m$.$^{10,14}$
We show that this 
quantization both with respect to the original action-angle variables
and the initial data
action-angle variables is equivalent.

In Section V, the classical holonomy operator of Section III
is quantized with respect to the initial data
action-angle variables.

The symbols $\rfloor$ and $\lfloor$ below stand for the left and right interior
products of multivector fields and exterior forms, respectively.

Let us recall that, given a fiber bundle $Y\to X$
coordinated by 
$(x^\la,y^i)$, a connection $K$ on $Y\to X$ is defined by a tangent-valued 
form
\be
K=dx^\la\ot(\dr_\la +k^i_\la\dr_i)
\ee
on $Y$.$^{15}$ A connection on a fiber bundle
$Y\to X$ is said to be an Ehresmann connection if, given an arbitrary smooth
curve $\xi([0,1])\subset X$, there exists its horizontal lift through any
point of $Y$ over $\xi(0)$.

Let $X$ be a real axis $\Bbb R$ provided with the Cartesian
coordinate $t$ 
possessing transition functions $t'=t+$const. A connection $K$ on a fiber
bundle $Y\to\Bbb R$ is uniquely represented by a vector field $K$ on $Y$ such
that $K\rfloor dt=1$.$^{12}$ This is the case of time-dependent mechanics.

\bigskip
\bigskip

\noindent
{\bf II. CLASSICAL TIME-DEPENDENT CIS}
\bigskip

Recall that the configuration space of time-dependent mechanics
is a fiber bundle $Q\to \Bbb R$
over the time axis $\Bbb R$. Let it be equipped with the bundle
coordinates $(t,q^k)$, $k=1,\ldots,m$.
The corresponding phase space is the vertical
cotangent bundle
$V^*Q$ of $Q\to\Bbb R$ endowed with the induced
coordinates $(t,q^k,p_k)$ relative to the holonomic coframes
$\{dq^k\}$.$^{12,16}$ 
The cotangent bundle
$T^*Q$ of $Q\to\Bbb R$ 
plays a role of the homogeneous phase space of time-dependent mechanics. 
It is equipped with the induced
coordinates $(t,q^k,p,p_k)$ relative to the holonomic coframes $\{dt,dq^k\}$.
With respect to this coordinates, the canonical symplectic form 
and the
corresponding Poisson bracket on $T^*Q$ read
\be
&&\Om=dp\w dt +dp_k\w dq^k, \\
&& \{f,f'\}_T =\dr_pf\dr_t f' -\dr_t
f\dr_p f'+ \dr^k f\dr_k f' -\dr_k
f\dr^k f', \qquad f,f'\in C^\infty(T^*Q). 
\ee
There is the one-dimensional trivial affine bundle
\mar{z11}\beq
\zeta:T^*Q\to V^*Q. \label{z11}
\eeq
As a consequence, the phase space
$V^*Q$ of time-dependent mechanics is provided with the canonical Poisson
structure
\mar{m72'}\beq
\{f,f'\}_V = \dr^kf\dr_kf'-\dr_kf\dr^kf',
\qquad  f,f'\in C^\infty(V^*Q), \label{m72'}
\eeq
given by the relations
\be
\zeta^*\{f,f'\}_V=\{\zeta^*f,\zeta^*f'\}_T, 
\qquad  
f,f'\in C^\infty(V^*Q). 
\ee
The corresponding Poisson bivector on $V^*Q$ reads $w_V=\dr_k\w \dr^k$.

A Hamiltonian of time-dependent mechanics is defined
as a global section
\mar{qqq}\beq
h: V^*Q\to T^*Q, \qquad  p\circ h=-\cH(t,q^j,p_j), \label{qqq}
\eeq
of the affine bundle $\zeta$ (\ref{z11}).$^{12,16}$ Given the 
pull-back form $h^*\Om$, the relations $\g_H\rfloor dt=1$, $\g_H\rfloor
h^*\Om=0$ 
define a unique Hamilton
vector field 
\mar{z3}\beq
\g_H=\dr_t + \dr^k\cH\dr_k- \dr_k\cH\dr^k \label{z3}
\eeq
on $V^*Q$ and the corresponding Hamilton equations
\mar{z20}\beq
\dot q^k=\dr^k\cH, \qquad \dot p_k=-\dr_k\cH. \label{z20}
\eeq

Note that, given a connection $\G=\dr_t +\G^i_t\dr_i$ 
on $Q\to\Bbb R$, 
any Hamiltonian $\cH$ (\ref{qqq}) admits the decomposition
$\cH=p_i\G^i_t +\wt\cH$
where $\wt\cH$ is a function on $V^*Q$.

An integral of motion of
the Hamilton equations (\ref{z20}) is a smooth real function $F$ 
on $V^*Q$ whose Lie derivative 
\be
\bL_{\g_H} F=\g_H\rfloor dF=\dr_tF +\{\cH,F\}_V 
\ee
along the Hamilton vector field $\g_H$ (\ref{z3}) vanishes. 
A time-dependent Hamiltonian system of $m$ degrees of freedom
is a CIS
if there exist $m$ independent integrals of motion
$\{F_k\}$ 
in involution with respect to the Poisson bracket $\{,\}_V$ (\ref{m72'}).
Their Hamiltonian vector fields 
\be
\vt_i=-w_V\lfloor dF_i =\dr^kF_i\dr_k - \dr_k F_i\dr^k 
\ee
and the Hamilton vector field $\g_H$ (\ref{z3}) 
generate a smooth regular distribution on the phase space $V^*Q$ 
and the corresponding foliation of $V^*Q$ in 
invariant manifolds. 

One can associate to any time-dependent CIS 
on $V^*Q$ an autonomous CIS on the homogeneous phase space $T^*Q$ as follows.

Given a Hamiltonian $h$ (\ref{qqq}), let us consider 
an autonomous Hamiltonian system on the symplectic
manifold $(T^*Q,\Om)$ with the Hamiltonian
\be
\bH=\dr_t\rfloor(\Xi-\zeta^* h^*\Xi)=p+\cH. 
\ee 
Its Hamiltonian vector field 
\mar{z5}\beq
\g_T=\dr_t -\dr_t\cH\dr_p+ \dr^k\cH\dr_k- \dr_k\cH\dr^k \label{z5}
\eeq
is projected onto the Hamilton vector field $\g_H$ (\ref{z3}) on $V^*Q$ so that
\be
\zeta^*(\bL_{\g_H}f)=\{\bH,\zeta^*f\}_T, \qquad f\in C^\infty(V^*Q).
\ee
An immediate consequence of this relation is the following.

(i) Given a time-dependent CIS $(H;F_k)$ on $V^*Q$, the 
Hamiltonian system $\{\bH,\zeta^*F_k\}$ on $T^*Q$ is a CIS.

(ii) If $M\subset V^*Q$ is an invariant manifold of the time-dependent
CIS $\{H;F_k\}$,  
then $h(M)\subset
T^*Q$ is an invariant manifold of 
the homogeneous CIS $(\bH,\zeta^*F_k)$.

Hereafter, let the Hamilton 
vector field $\g_H$ (\ref{z3}) be complete, i.e., the Hamilton equations
(\ref{z20}) admit a unique global solution (a trajectory of $\g_H$)
through every point of the
 phase space $V^*Q$. The trajectories of $\g_H$ define a trivial bundle 
$V^*Q\to V^*_0Q$
over the fiber $V^*_0Q$
of $V^*Q\to\Bbb R$ at $t=0$. Then any invariant manifold 
$M$ of $\{H;F_k\}$ is also a trivial bundle
$M= \Bbb R\times M_0$ over $M_0=M\cap V^*_0Q$.

If $M_0$ is compact, one can
introduce   
action-angle coordinates around an
invariant manifold $M$ 
by use of the action-angle
coordinates around the invariant manifold $h(M)$ of the
corresponding autonomous CIS on $T^*Q$.$^{10}$
Namely, $h(M)$ has an open
neighbourhood which is a trivial bundle
\mar{z41}\beq
U'=V'\times\Bbb R\times T^m\to V'\times\Bbb R\to V' \label{z41}
\eeq
over a domain $V'\subset \Bbb R^{m+1}$
with respect to the action-angle coordinates 
$(I_0,I_i,t,\f^i)$. Herewith, the following holds.
(i) 
$I_0=\bH$. (ii) The integrals of motion $\zeta^*F_k$
depend only on the action coordinates $I_i$. (iii)
The symplectic form $\Om$ on 
$U'$ reads
\be
\Om=dI_0\w dt + dI_i\w d\f^i. 
\ee
The symplectic annulus $U'$ (\ref{z41}) inherits the fibration structure
(\ref{z11}) over the toroidal domain
\mar{z46}\beq
 U=V\times \Bbb R\times T^m, \qquad V\subset \Bbb R^m. \label{z46}
\eeq
Coordinated by $(I_i,t,\f^i)$
and provided with the Poisson structure (\ref{m72'}), the toroidal  
domain (\ref{z46}) is a
phase space of the time-dependent CIS $(H;F_i)$
around its instantly compact invariant manifold $M$. 
Since $\bH=I_0$, the Hamilton vector field (\ref{z5})
 is $\g_T=\dr_t$, and so is its projection $\g_H$ (\ref{z3})
onto $U$. Hence, the above-mentioned action-angle coordinates $(I_i,t,\f^i)$ 
are the initial data coordinates.

These action-angle coordinates are by no means unique. Let $\cH$ be an
arbitrary smooth function on $\Bbb R^m$. Then the canonical transformation
\mar{r5}\beq
I'_0=I_0-\cH(I_j), \qquad I'_i=I_i, \qquad t'=t, \qquad \vf^i= \f^i +t\dr^i\cH(I_j) 
\label{r5}
\eeq
gives new action-angle coordinates corresponding to a different trivialization
of $U'$ (\ref{z41}) (and $U$ (\ref{z46})).
Accordingly, the Hamilton vector field $\g_H$ 
takes the form (\ref{z3}), and the Hamilton equations (\ref{z20}) read
\be
\dot \vf^k=\dr^k\cH, \qquad \dot I_k=0. 
\ee
These are the Hamilton equations of an autonomous CIS with a time-independent
Hamiltonian $\cH$ on the toroidal 
domain $U$ (\ref{z46}). 
\bigskip
\bigskip

\noindent
{\bf III. CLASSICAL HOLONOMY OPERATORS}
\bigskip

The phase space of a Hamiltonian system
with time-dependent parameters is a composite fiber bundle
$\Pi\to\Si\times \Bbb R\to\Bbb R$,
where $\Pi\to\Si\times\Bbb R$ is a symplectic bundle 
and $\Si\times \Bbb R\to\Bbb R$ is a parameter
bundle whose sections are parameter functions.$^{12,13,17,18}$
In the case under consideration, all bundles are
trivial and their trivializations hold fixed. 
Namely, the phase space is the product
\be
\Pi= \Si\times U=\Si\times (V\times \Bbb R\times T^m)\to \Si\times\Bbb
R\to\Bbb R,  
\ee
equipped with the coordinates $(\si^\al,I_k,t,\f^k)$. Let us
suppose for a time that parameters are also dynamic variables. The
 phase space of this system is the fiber bundle
\be
\Pi'=T^*\Si\times U\to \Si\times\Bbb R\times T^m
\ee
coordinated by $(\si^\al,\si_\al,I_k,t,\f^k)$.
A generic Hamiltonian of such a system is 
\mar{pr30}\beq
 \cH_\Si=\si_\al\Si^\al_t +I_k(\La^k_t +\La^k_\al\Si^\al_t) 
+\wt\cH(\si^\bt, I_j,t,\f^j), \label{pr30}
\eeq
where 
\be
\dr_t +\Si^\al_t\dr_\al +(\La^k_t +\La^k_\al\Si^\al_t)\dr_k
\ee
is a composite connection on the fiber bundle 
$\Si\times\Bbb R\times T^m\to \Bbb R$
generated by a connection $\dr_t +\Si^\al_t\dr_\al$  
on the parameter bundle $\Si\times\Bbb R\to\Bbb R$ and a connection
\mar{r30}\beq
\La=dt\ot(\dr_t +\La^k_t\dr_k) + d\si^\al\ot(\dr_\al +\La^k_\al\dr_k) \label{r30}
\eeq
on $\Si\times\Bbb R\times T^m\to \Si\times\Bbb R$.$^{12,13,18}$ 
Then a Hamiltonian system with a fixed
parameter function  $\si^\al=\xi^\al(t)$ is characterized by 
the Hamiltonian 
\mar{r12}\beq
\cH_\xi=I_k[\La^k_t(t,\f^j)+ \La^k_\al
(\xi^\bt,t,\f^j)\dr_t\xi^\al]
+\wt\cH(\xi^\bt,I_j,t,\f^j) \label{r12}
\eeq
on the pull-back bundle $U=\xi^*\Pi$ (\ref{z46}). 

Let $(I_k,t,\f^k)$ be the initial data action-angle coordinates of a
time-dependent CIS. Its Hamiltonian $\cH$ with respect to these coordinates
vanishes. Therefore, we can introduce a desired
holonomy operator by the appropriate choice of the connection
$\La$ (\ref{r30}). 
Let us put $\La^k_t=0$ and 
assume that coefficients $\La^k_\al$ 
are independent 
of time, i.e., the
part 
\mar{r31}\beq
\La_\Si=d\si^\al\ot(\dr_\al +\La^k_\al\dr_k) \label{r31}
\eeq
of the connection $\La$ (\ref{r30})
is a connection on the fiber bundle (\ref{r25}).
Then the Hamiltonian of a perturbed CIS reads
\mar{r14}\beq
\cH_\xi= I_k \La^k_\al
(\xi^\bt,\f^j)\dr_t\xi^\al. \label{r14}
\eeq
Its Hamilton vector field (\ref{z3}) is
\mar{r21}\beq
\g_H=\dr_t + \La^i_\al \dr_t\xi^\al\dr_i
-I_k\dr_i\La^k_\al \dr_t\xi^\al\dr^i. \label{r21} 
\eeq
It leads to the Hamilton equations
\mar{zz10,1}\ben
&& \dr_t\f^i=\La^i_\al(\xi(t),\f^l) \dr_t\xi^\al, \label{zz10}\\
&& \dr_t I_i=-I_k\dr_i\La^k_\al(\xi(t),\f^l) \dr_t\xi^\al. \label{zz11}
\een
Note that
\mar{r22}\beq
V^*\La_\Si=d\si^\al\ot(\dr_\al +\La^i_\al\dr_i -I_k\dr_i\La^k_\al\dr^i) \label{r22}
\eeq
is the lift  of the connection $\La_\Si$ (\ref{r31}) onto the fiber bundle 
$\Si\times (V\times T^m)\to \Si$, seen as a subbundle of 
the vertical cotangent bundle $V^*(\Si\times T^m)=\Si\times T^*T^m$
of the fiber bundle (\ref{r25}). It follows that any solution $I_i(t)$, $\f^i(t)$
of the Hamilton 
equations (\ref{zz10}) -- (\ref{zz11}) (i.e., an integral curve of the 
Hamilton vector field (\ref{r21})) is a horizontal lift  of the curve
$\xi(t)\subset \Si$ with respect to the connection $V^*\La_\Si$ (\ref{r22}).
i.e., $I_i(t)=I_i(\xi(t))$, $\f^i(t)=\f(\xi(t))$. 
Thus, the right-hand side of the Hamilton equations (\ref{zz10}) -- (\ref{zz11}) is the 
holonomy operator
\mar{r33}\beq
\Delta= (\La^i_\al \dr_t\xi^\al, -I_k\dr_i\La^k_\al \dr_t\xi^\al) \label{r33}
\eeq
(cf. (\ref{r0}) where $\wh S_0=0$). It is not a linear operator, but
the substitution of a solution
$\f(\xi(t))$ of the equation (\ref{zz10}) into the Hamilton equation
(\ref{zz11})  
results in a linear holonomy operator on 
the action variables $I_i$.

Let us show that the holonomy operator (\ref{r33}) is well defined. 
Since any vector field $\vt$
on $\Bbb R\times T^m$ such that $\vt\rfloor dt=1$ is complete, the
Hamilton equation 
(\ref{zz10}) has solutions for any parameter function $\xi(t)$. It
follows that  
any connection $\La_\Si$ (\ref{r31}) on the fiber bundle
(\ref{r25}) is an Ehresmann connection, and so is its lift
(\ref{r22}). Therefore,
any curve $\xi([0,1])\subset \Si$ can play the role of the parameter function
in the holonomy operator $\Delta$ (\ref{r33}).

Now, let us return to the original action-angle coordinates
$(I_k,t,\vf^k)$ by means of the canonical transformation (\ref{r5}).
Relative to these coordinates, the perturbed Hamiltonian reads
\be
\cH'= I_k\La^k_\al(\xi(t), \vf^i-t\dr^i\cH(I_j))\dr_t\xi^\al(t) + \cH(I_j),
\ee
and the Hamilton equations (\ref{zz10}) -- (\ref{zz11}) take the form
\be
&& \dr_t\vf^i=\dr^i\cH(I_j)+ \La^i_\al(\xi(t), \vf^l-t\dr^l\cH(I_j))
\dr_t\xi^\al(t) \\ 
&& \qquad - tI_k\dr^i\dr^s\cH(I_j)\dr_s\La^k_\al(\xi(t),
\vf^l-t\dr^l\cH(I_j))\dr_t\xi^\al(t),\\ 
&& \dr_t I_i=-I_k\dr_i\La^k_\al(\xi(t), \vf^l-t\dr^l\cH(I_j))
\dr_t\xi^\al(t).   
\ee
Their solution is $I_i(\xi(t))$, $\vf^i(t)=\f^i(\xi(t))
+t\dr^i\cH(I_j(\xi(t)))$ where 
$I_i(\xi(t))$, $\f^i(\xi(t))$ is a solution 
of the Hamilton equations (\ref{zz10}) -- (\ref{zz11}). 
It is readily observed that the
action variables $I_k$ are driven only by the holonomy operator, while
the angle 
variables $\vf^i$ have a nongeometric summand.
 
Let us emphasize that, in the construction of the holonomy operator
(\ref{r33}), we did not impose  
any restriction on the connection
$\La_\Si$ (\ref{r31}). Therefore, any connection on the fiber
bundle (\ref{r25}) generates a holonomy operator in a CIS. However, 
a glance at the expression (\ref{r33}) shows that this operator
becomes zero on action variables if all coefficients $\La^k_\la$ of the
connection $\La_\Si$ (\ref{r31}) are constant, i.e., $\La_\Si$ 
is a principal 
connection on the fiber bundle (\ref{r25}) seen as a principal bundle with 
the structure group $T^m$.

\bigskip
\bigskip

\noindent
{\bf IV. QUANTUM CIS}
\bigskip

There are different approaches to quantization of CISs.$^{19}$ 
Their geometric quantization was studied at first
with respect to the polarization spanned by Hamiltonian vector fields of
integrals of motion.$^{20}$ For example, the well-known 
Simms quantization of
the harmonic oscillator is of this type.
In this approach, the problem is that the associated quantum algebra
includes affine functions of angle coordinates which are ill defined.
As a consequence, elements of the carrier space of this quantization
fail to be smooth, 
but are tempered distributions. In recent works,$^{10,14}$
we have developed a different variant of geometric quantization
of CISs by use of the angle polarization
spanned by almost-Hamiltonian vector fields $\dr^k$ of angle variables. 
This quantization is equivalent to geometric quantization of the cotangent 
bundle $T^*T^m$ of a torus $T^m$ with respect to the vertical
polarization. The result is as follows.

Given an autonomous CIS on a symplectic annulus
\be
P=V\times T^m, \qquad \Om_P=dI_i\w d\vf^i 
\ee
equipped with the action-angle coordinates $(I_i,\vf^i)$,
its quantum algebra $\cA$ with respect to the above mentioned 
angle polarization 
consists of affine functions
\be
f=a^k(\vf^j)I_k + b(\vf^j) 
\ee
of action coordinates $I_k$. They are represented by 
self-adjoint unbounded operators
\mar{lmp135}\beq
\wh f=
-ia^k\dr_k-\frac{i}{2}\dr_ka^k-a^k\la_k +
b \label{lmp135}
\eeq
in the separable pre-Hilbert
space  of complex half-forms 
on $T^m$. If coordinate transformations of $T^m$ 
are only translations, this
space can be identified with the pre-Hilbert space 
$\Bbb C^\infty(T^m)$
of smooth complex functions on $T^m$.
Different tuples of real numbers $(\la_1,\ldots,\la_m)$ and
$(\la'_1,\ldots,\la'_m)$ specify inequivalent representations
(\ref{lmp135}), unless $\la_k-\la'_k\in\Bbb Z$ for all $k=1,\ldots,m$.
These numbers come from the de Rham cohomology group 
$H^1(T^m)=\Bbb R^m$. 

In particular, the action operators (\ref{lmp135}) read
$\wh I_k=-i\dr_k -\la_k$. They are bounded.
By virtue of the multidimensional Fourier theorem, 
an orthonormal basis for $\Bbb C^\infty(T^m)$ consists of
functions
\mar{ci15}\beq
\psi_{(n_r)}(\vf)=\exp[in_r\vf^r], \qquad (n_r)=(n_1,\ldots,n_m)\in\Bbb Z^m. 
\label{ci15}
\eeq
With respect to this basis, the action operators are 
brought into countable diagonal matrices
\mar{ci9}\beq
\wh I_k\psi_{(n_r)}=(n_k-\la_k)\psi_{(n_r)}, \label{ci9}
\eeq
while functions $a^k(\vf)$ are decomposed in Fourier series
of the functions $\psi_{(n_r)}$,
which act on $\Bbb C^\infty(T^m)$ by the law
\mar{ci11}\beq
\wh\psi_{(n_r)} \psi_{(n'_r)}=\psi_{(n_r+n'_r)}. \label{ci11}
\eeq
It should be emphasized that $\wh a^k\wh I_k\neq \wh{a^kI_k}\neq
\wh{I_ka^k}$.

If a Hamiltonian $\cH(I_k)$ of
an autonomous CIS is an analytic function on $\Bbb R^m$,
it is uniquely quantized as a Hermitian element
$\wh\cH(I_k)=\cH(\wh I_k)$ of the enveloping algebra
 of $\cA$. It is a bounded self-adjoint
operator with the countable spectrum
\mar{ww10}\beq
\wh \cH(I_k)\psi_{(n_r)}=E_{(n_r)}\psi_{(n_r)}, \qquad 
E_{(n_r)}=\cH(n_k-\la_k), \qquad n_k \in(n_r). \label{ww10}
\eeq

In order to quantize a time-dependent CIS on the Poisson toroidal domain $(U,\{,\}_V)$
(\ref{z46}) equipped with action-angle coordinates
$(I_i,t,\vf^i)$, one may follow the instantwise
geometric quantization of time-dependent mechanics.$^{21}$ 
As a result, we can simply replace functions on $T^m$ with those on $\Bbb R\times
T^m$.$^{10}$ Namely, 
the corresponding quantum algebra $\cA\subset C^\infty(U)$ consists of
affine functions
\mar{ww7}\beq
f=a^k(t,\vf^j)I_k + b(t,\vf^j)  \label{ww7}
\eeq 
of action coordinates $I_k$ represented by the operators (\ref{lmp135})
in the space 
\mar{r41}\beq
E=\Bbb C^\infty(\Bbb R\times T^m) \label{r41}
\eeq
of smooth complex functions $\psi(t,\vf)$ on
$\Bbb R\times T^m$. This space is provided with the structure of the
pre-Hilbert $\Bbb C^\infty(\Bbb R)$-module with respect to the nondegenerate 
$\Bbb C^\infty(\Bbb R)$-bilinear form
\be
\lng \psi|\psi'\rng=\left(\frac1{2\pi}\right)^m\op\int_{T^m} \psi \ol \psi'd^m\vf, 
\qquad \psi,\psi'\in \Bbb C^\infty(\Bbb R\times T^m).  
\ee 
Its basis consists of the pull-back onto
$\Bbb R\times T^m$ of the 
functions $\psi_{(n_r)}$ (\ref{ci15}). 

This quantization
of a time-dependent CIS is extended to 
the associated homogeneous CIS on the symplectic annulus $(U',\Om)$ (\ref{z41})
by means of the
operator $\wh I_0=-i\dr_t$ in the pre-Hilbert module $E$ (\ref{r41}). 
Accordingly, the homogeneous Hamiltonian $\bH$ 
is quantized as $\wh\bH=-i\dr_t +\wh\cH$. The corresponding
Schr\"odinger equation is 
\mar{r42}\beq
\wh\bH\psi=-i\dr_t\psi +\wh\cH\psi=0, \qquad 
\psi\in  E. \label{r42}
\eeq

For instance, the quantum Hamiltonian of the original autonomous CIS is
\be
\wh\bH=-i\dr_t +\cH(\wh I_j). 
\ee  
Its spectrum $\wh \bH\psi_{(n_r)}=E_{(n_r)}\psi_{(n_r)}$ relative to
the basis   $\{\psi_{(n_r)}\}$ for 
$E$ (\ref{r41}) coincides with that  of 
the autonomous Hamiltonian (\ref{ww10}). The Schr\"odinger equation (\ref{r42})
reads
\be
\wh\bH\psi=-i\dr_t\psi +\cH(-i\dr_k -\la_k)\psi=0, \qquad 
\psi\in E. 
\ee
Its solutions are the Fourier series
\be
\psi=\op\sum_{(n_r)} B_{(n_r)} \exp[-itE_{(n_r)}]\psi_{(n_r)}, \qquad 
B_{(n_r)}\in\Bbb C.
\ee

Now, let us quantize this CIS with respect to the initial data 
action-angle coordinates $(I_i,\f^i)$. Its quantum algebra
$\cA_0\subset C^\infty(U)$ consists of affine functions 
\mar{r45}\beq
f=a^k(t,\f^j)I_k + b(t,\f^j). \label{r45}
\eeq
The canonical transformation (\ref{r5}) provides an isomorphism between
Poisson algebras 
$\cA$ and $\cA_0$. Functions $f$ (\ref{r45}) are represented by
the operators
$\wh f$ (\ref{lmp135}) in the pre-Hilbert module $E_0$ 
of smooth complex functions $\Psi(t,\f)$ on
$\Bbb R\times T^m$. Given its
basis $\Psi_{(n_r)}(\f)=[in_r\f^r]$, the operators $\wh I_k$ and $\wh\psi_{(n_r)}$
take the form (\ref{ci9}) and (\ref{ci11}), respectively. The Hamiltonian of 
a quantum CIS with respect to the initial data variables is $\wh\bH_0=-i\dr_t$.
Then one easily obtains the isometric isomorphism
\mar{r46}\beq
R(\psi_{(n_r)})=\exp[itE_{(n_r)}]\Psi_{(n_r)}, \qquad \lng R(\psi)| R(\psi')\rng=
\lng \psi | \psi'\rng, \label{r46}
\eeq
between the pre-Hilbert modules $E$ and $E_0$ which provides the equivalence
\mar{r47}\beq
\wh I_i=R^{-1}\wh I_iR, \qquad \wh\psi_{(n_r)}=R^{-1}\wh\Psi_{(n_r)} R,
\qquad \wh\bH= R^{-1}\wh\bH_0 R \label{r47}
\eeq 
of the quantizations of a CIS with respect to the original and initial data
action-angle variables.
\bigskip
\bigskip

\noindent
{\bf V. QUANTUM HOLONOMY OPERATORS}
\bigskip

In view of the isomorphism (\ref{r47}), let us first construct
a holonomy operator for a quantum CIS $(\cA_0,\wh\bH_0)$
with respect to the initial data action-angle 
coordinates. Let us consider the perturbed homogeneous Hamiltonian
\be
\bH_\xi=\bH_0+\bH_1=I_0+\dr_t\xi^\al(t)\La_\al^k(\xi(t),\f^j)I_k 
\ee
of the classical perturbed system (\ref{r14}). 
Its perturbation term $\bH_1$ is of the form (\ref{ww7}) and, therefore,
is quantized by the operator
\be
\wh\bH_1=-i\dr_t\xi^\al\wh\Delta_\al=-i\dr_t\xi^\al[\La^k_\al\dr_k 
+\frac12\dr_k(\La^k_\al) -i\la_k \La^k_\al].
\ee

The quantum Hamiltonian $\wh\bH_\xi=\wh \bH_0+\wh\bH_1$ defines 
the Schr\"odinger equation  
\mar{r52}\beq
\dr_t\Psi+ \dr_t\xi^\al[\La^k_\al\dr_k 
+\frac12\dr_k(\La^k_\al) -i\la_k \La^k_\al]\Psi=0. \label{r52}
\eeq
If a solution exists, it can be written by means of the evolution
operator which reduces to the geometric factor $U_1$ (\ref{r4}).
The latter can be viewed as a displacement operator along the curve 
$\xi[0,1]\subset \Si$ with respect to the connection
\mar{r53}\beq
\wh\La_\Si=d\si^\al(\dr_\al +\wh\Delta_\al) \label{r53}
\eeq 
in the $\Bbb C^\infty(\Si)$-module $\Bbb C^\infty(\Si\times T^m)$ 
of smooth complex functions on $\Si\times T^m$.$^{13,15,18,22}$
Let us study the existence if this displacement operator.

Given a connection $\La_\Si$ (\ref{r31}), let $\Phi^i(t,\f)$ 
denote the flow of the
complete vector field $\dr_t+\La_\al^i(\xi,\f)\dr_t\xi^\al\partial_i$ 
on $\Bbb R\times T^m$. It is a solution  
of the Hamilton equation (\ref{zz10}) with the initial data $\f$. 
We need the inverse flow $(\Phi^{-1})^i(t,\f)$ which obeys the equation
\be
\dr_t(\Phi^{-1})^i(t,\f)=-\dr_t\xi^\al\La^i_\al(\xi,(\Phi^{-1})^i(t,\f))=
-\dr_t\xi^\al\La^k_\al(\xi,\f)\dr_k(\Phi^{-1})^i(t,\f).
\ee
Let $\Psi_0$ be an arbitrary
complex half-form $\Psi_0$ on $T^m$ possessing identical transition functions,
and let the same symbol stand for its pull-back
onto $\Bbb R\times T^m$. Given its pull-back
\mar{r56}\beq
(\Phi^{-1})^*\Psi_0=
\det\left(\frac{\dr (\Phi^{-1})^i}{\dr \f^k}\right)^{1/2}\Psi_0(\Phi^{-1}(t,\f)),
\label{r56}
\eeq
it is readily observed that
\be
\Psi=(\Phi^{-1})^*\Psi_0\exp[i\la_k\f^k] 
\ee
obeys the Schr\"odinger equation (\ref{r52}) with the initial data $\Psi_0$. 
This function is well defined only if all the numbers 
$\la_k$ equal $0$ or $\pm 1/2$.
Note that, if some numbers $\la_k$ are equal to $\pm 1/2$, then 
$\Psi_0\exp[i\la_k\f^k]$ 
is a half-density on $T^m$ whose transition functions equal $\pm 1$, i.e., 
it is a section of a nontrivial metalinear bundle over $T^m$ 

We thus observe that if $\la_k$ equal $0$ or $\pm 1/2$, then the displacement
operator always exists and $\Delta=i\bH_1$ is a holonomy
operator. A glance at the action law (\ref{ci11}) shows that this
operator is infinite-dimensional.  

For instance, let $\La_\Si$ (\ref{r31}) be the above mentioned principal 
connection, i.e., $\La^k_\al=$const. Then the Schr\"odinger equation
(\ref{r52}) where $\la_k=0$ takes the form
\be
\dr_t\Psi(t,\f^j)+ \dr_t\xi^\al(t)\La^k_\al\dr_k\Psi(t,\f^j)=0.
\ee
Its solution (\ref{r56}) is 
\be
\Psi(t,\f^j)=\Psi_0(\f^j-(\xi^\al(t)-\xi^\al(0))\La^j_\al).
\ee
The corresponding evolution operator reduces to Berry's phase multiplier
\be
U_1\Psi_{(n_r)}= \exp[-in_j(\xi^\al(t)-\xi^\al(0))\La^j_\al)]\Psi_{(n_r)}, 
\qquad n_j\in (n_r).
\ee
It keeps the eigenvectors of the action operators $\wh I_i$.

In order to return to the original action-angle variables, one can employ the 
morphism $R$ (\ref{r46}). The corresponding Hamiltonian reads 
$\bH'=R^{-1}\bH_\xi R$. The key point is that, due to the relation (\ref{r47}),
the action operators  $\wh I_i$ have the same mean values
\be
\lng I_k\psi| \psi\rng= \lng I_k\Psi | \Psi\rng, \qquad \Psi=R(\psi),
\ee
with respect both to the original and the initial data action-angle
variables. Therefore, these mean values are defined only by the holonomy
operator.  

\bigskip
\bigskip

\noindent
{\bf VI. CONCLUSIONS}
\bigskip

We have shown that any CIS around its compact invariant manifold admits
a perturbation dependent on parameters by means of holonomy operator
associated to a connection on the fiber bundle (\ref{r25}).

Since action variables are driven only by 
a holonomy operator, one can use this operator 
in order to perform  a dynamic transition 
between classical solutions or quantum states of 
an unperturbed CIS
by an appropriate choice of a parameter function $\xi$. 
The key point is that 
this transition can take an arbitrary short time 
because we are entirely free 
with time parametrization of $\xi$ and can choose it quickly changing,
in contrast with slowly varying parameter functions in adiabatic models. 
For instance, one can choose $\xi$ a step function, then its time
derivative is a $\dl$-function of time.
This fact makes 
nonadiabatic holonomy operators in CISs promising for 
several applications, including classical and quantum scattering in 
integrable Hamiltonian systems,$^{23}$
quantum control operators,$^{24,25}$ and the above mentioned quantum
computation.  
It also looks attractive that quantum 
holonomy operators in CISs are essentially 
infinite-dimensional, whereas
both the existent quantum control theory and
the theory of quantum information and computation$^{26}$
involve only finite-dimensional operators.


\begin{thebibliography}{ddd}

\bibitem{zan} P.Zanardi and M.Rasetti, {\it Phys. Lett. A} {\bf
264}, 94 (1999).
 
\bibitem{fuj} K.Fujii, {\it J. Math. Phys.} {\bf 41}, 4406 (2000).

\bibitem{pach} J.Pachos and P.Zanardi, {\it Int. J. Mod. Phys.} {\bf B15},
1257 (2001).

\bibitem{anan} Y.Aharonov and J.Anandan, {\it Phys. Rev. Lett.} {\bf 58},
1593 (1987).

\bibitem{bohm} A.Bohm and A.Mostafazadeh, {\it J. Math. Phys.}
{\bf 35}, 1463 (1994).

\bibitem{xian} W.Xiangbin and M.Keiji, {\it J. Phys. A} {\bf 34}, L631 (2001).

\bibitem{well} C.Wellard, L.C.L.Hollenberg and H.C.Pauli, {\it Phys.
Rev. A} {\bf 65}, 032303 (2002).

\bibitem{zhu} Shi-Liang Zhu and Z.D.Wang, {\it Phys. Rev. Lett.} {\bf
89}, 097902 (2002). 

\bibitem{dew} A.Dewisme and S.Bouquet, {\it J. Math. Phys} {\bf 34},
997 (1993).

\bibitem{acang1} F.Fiorani, G.Giachetta and G.Sardanashvily,
{\it J. Math. Phys.} {\bf 43}, 5013 (2002).

\bibitem{acang2} G.Giachetta, L.Mangiarotti and G.Sardanashvily, 
{\it J. Phys. A} {\bf 35}, L439 (2002).

\bibitem{book98} L.Mangiarotti and G.Sardanashvily, {\it Gauge Mechanics}
(World Scientific, Singapore, 1998).

\bibitem{jmp00b} G.Sardanashvily, {\it J. Math. Phys.} {\bf 41}, 5245 (2000).

\bibitem{phlet} G.Giachetta, L.Mangiarotti and G.Sardanashvily, 
{\it Phys. Lett. A} {\bf 301}, 53 (2002).


\bibitem{book00} L.Mangiarotti and G.Sardanashvily, {\it Connections in
Classical and Quantum Field Theory} (World Scientific, Singapore, 2000)

\bibitem{sard98} G.Sardanashvily,
{\it J. Math. Phys.} {\bf 39}, 2714 (1998).

\bibitem{wu} Y.Wu, {\it J. Math. Phys.} {\bf 31}, 294 (1990).

\bibitem{jmp02} G.Giachetta, L.Mangiarotti 
and G.Sardanashvily, 
{\it J. Math. Phys.} {\bf 43}, 2882 (2002).

\bibitem{gos} M.Gosson, {\it J. Phys. A} {\bf 34}, 10085 (2001).

\bibitem{myk} I.Mykytiuk,
A.Prykarpatsky, R.Andrushkiw and V.Samoilenko, {\it J. Math. Phys.}
{\bf 35}, 1532 (1994).

\bibitem{sard02} G.Giachetta, L.Mangiarotti and G.Sardanashvily, 
{\it J. Math. Phys.} {\bf 43}, 56 (2002).

\bibitem{iliev} B.Iliev, {\it J. Phys. A} {\bf 34}, 4887 (2001).

\bibitem{jung} C.Jung and T.Seligman, {\it Phys. Rep.} {\bf 285}, 77 (1997).

\bibitem{dal} D.D'Alessandro, {\it J. Math. Phys.} {\bf 42}, 4488 (2001).

\bibitem{sch} S.Schirmer, I.Pullen and A.Solomon, {\it J. Phys. A} {\bf
35}, 2327 (2002).

\bibitem{keyl} M.Keyl, {\it Phys. Rep.} {\bf 369}, 431 (2002).

   

\end{thebibliography}
\end{document}